**Direct vs. Indirect Measurement of the Effective Electronic Temperature in Quantum Dot Solids**

Anton Kompatscher[+], Morteza Shokrani[+], Johanna Feurstein, Martijn Kemerink[*]

Institute for Molecular Systems Engineering and Advanced Materials, Heidelberg University, Im Neuenheimer Feld 225, 69120 Heidelberg, Germany

[+] contributed equally

[*] corresponding author; email: martijn.kemerink@uni-heidelberg.de

**Abstract**

One of the characteristics of disordered semiconductors is the slow thermalization of charge carriers after excitation due to photoabsorption or high electric fields. An elegant way to capture the effects of the latter on the conductivity is through a field-dependent effective electronic temperature $T_{eff}$ that can significantly exceed that of the lattice. Despite its elegance, its actual use has been limited, which, at least in part, can be attributed to the concept originating from computer simulations; experimental confirmations have largely been indirect (through scaling of conductivity) and did not establish that $T_{eff}$ equals the real temperature of the electron distribution. Moreover, it has hardly been tested for important classes of disordered materials, including quantum dot solids. Here, we investigate whether the effective temperature concept is applicable to quantum dot solids, using zinc oxide as relevant model system. To verify that field-driven conductivity increases indeed reflect an actual increase of the electronic temperature, we combine direct and indirect measurements of $T_{eff}$: we convert conductivity changes at high fields to an effective temperature that we show to be consistent with a direct measurement of the electronic temperature using the Seebeck effect. These results not only confirm the relevance of the effective temperature concept to quantum dot solids but also confirm its general physical reality and open the way to systematic investigations into charge carrier (de)localization in disordered media.

**Keywords:**

**Introduction**

Energetically disordered systems have strongly distinct properties from their more ordered counterparts, making them a fascinating subject of study. The description of conduction processes in them is typically described by hopping between localized sites in a strongly energy-dependent density of states (DOS). Here, we are specifically interested in the behavior at high electric fields. This is commonly modeled by stochastic models, including kinetic Monte-Carlo simulations, that predict a relatively slow thermalization of high-energy electrons (or holes) in such systems [1]. Specifically, Marianer and Shklovskii [2] predicted that under the application of a high electric field, the distribution of electron energies, i.e., the density of occupied states (DOOS), would shift out of thermal equilibrium. This new distribution was described in terms of an elevated 'effective' temperature $T_{eff}$ of the charge carriers in the unchanged DOS. Figure 1a schematically illustrates the concept. The physical rationale is that application of a high electric field provides the charge carriers with additional energy and hence changes their distribution in a way that mimics their behavior at higher lattice temperatures. Later theoretical studies extended the concept from exponential to gaussian densities of states, suggesting that the precise shape of the DOS is not critical, provided it is sufficiently (exponential or stronger) energy-dependent [3].

The most used mathematical description of the field-dependent effective temperature has been introduced by Marianer and Shklovskii [2], phenomenologically modelling it as:

$$T_{eff} = \left[ T_{lat}^{\beta} + \left( \gamma \frac{\alpha_l e |F|}{k_B} \right)^{\beta} \right]^{\frac{1}{\beta}} \qquad (1)$$

The formula includes the lattice temperature $T_{lat}$, the electric field $F$ as well as the wavefunction localization length $\alpha_l$. The parameters $\gamma$ and $\beta$ are typically used to adjust the model to either numerical simulations or experiments and are usually in the range of 0.6-0.9 and 1.5-2, respectively. To fulfill physical consistency, the parameter $\beta$ has to be larger than 1; we employ $\beta = 2$ and $\gamma = 0.89$. [4]

The effective temperature concept has been tested on hydrogenated silicon via both photoexcitation and conductivity experiments with mixed results. The conductivity experiments first conducted by Nebel et al. [5] are reasonably well described by this concept, i.e., eq. 1. Later photoconductivity experiments analyzing the phenomenon painted a mixed picture [6–10]. The concept was later also remarked on in the context of carbides [11,12]. There, conductivity scaling could well be described by an adapted formalism by dividing the data into low and high field ranges. The time-of-flight experiment on the other hand could not fully be explained by the theory since the effect was stronger than predicted. Viji et al. [13] show above-lattice electronic temperatures in organic solar cells via noise spectroscopy and linked this to a similar shift in electron and hole distributions due to photoexcitation. Markvart similarly proposes a link between hot carrier solar cells and thermoelectrics [14]. In recent work, we proposed a new method to directly measure the effective electronic temperature, based on the detection of the resulting Seebeck voltage and applied it to organic semiconductors. [15]

Summarizing, the relatively limited set of existing experiments are not very comparable and thus hard to check against each other; specifically, they do not allow to draw conclusions regarding the consistency of indirectly (using conductivity scaling) measured $T_{eff}$ and the actual temperature of the electronic distribution. Here, we aim to address this shortcoming by comparing the (direct) Seebeck- and the (indirect) conductivity-based experiments in the same material under identical conditions. If successful, this would validate the use of eq. 1 as a direct probe of charge carrier (de)localization in disordered semiconductors. We note that even though hopping rates typically depend exponentially on the wavefunction localization length $\alpha_l$, there have been few, if any, systematic experimental studies in that direction.

The indirect conductivity experiment measures the field and temperature dependent conductivity using well-defined structures as shown in fig. 1b. It then relates conductivity at low lattice temperatures $T_{lat}$ and high applied fields $F$ to the low-field (Ohmic) conductivity at elevated temperatures[5], assuming equal conductivity scaling with actual and effective temperature,

$$\sigma(T_{lat}, F) = \sigma(T_{eff}(T_{lat}, F), 0). \quad (2)$$

The direct, Seebeck-based, experiment (fig. 1c) creates spatially separated areas with high and low electric fields, inducing a gradient in the (effective) electronic temperature. The resulting Seebeck voltage is then measured and the effective temperature is calculated by knowing the Seebeck coefficient of the material, using $\Delta T_{eff} = -\Delta V/S$. The experiment therefore must consist of three terminals at minimum. Between two terminals, a driving voltage is applied to locally induce an enhanced effective temperature while the third terminal, that sits far from the others in a region of lower effective temperature, measures the resulting Seebeck voltage. Here, we can use the fact that the effective temperature only depends on the absolute value of the field and not its direction. Therefore, under the application of an AC driving signal, the Seebeck response has double the frequency of the driving signal, allowing us to separate it from the original input and most other spurious signals. A full discussion of the method can be found on our previous work Ref. [15]. We call this method direct since, in contrast to the conductivity-scaling-based experiment, (electronic) temperature is the only factor causing the measured voltage.

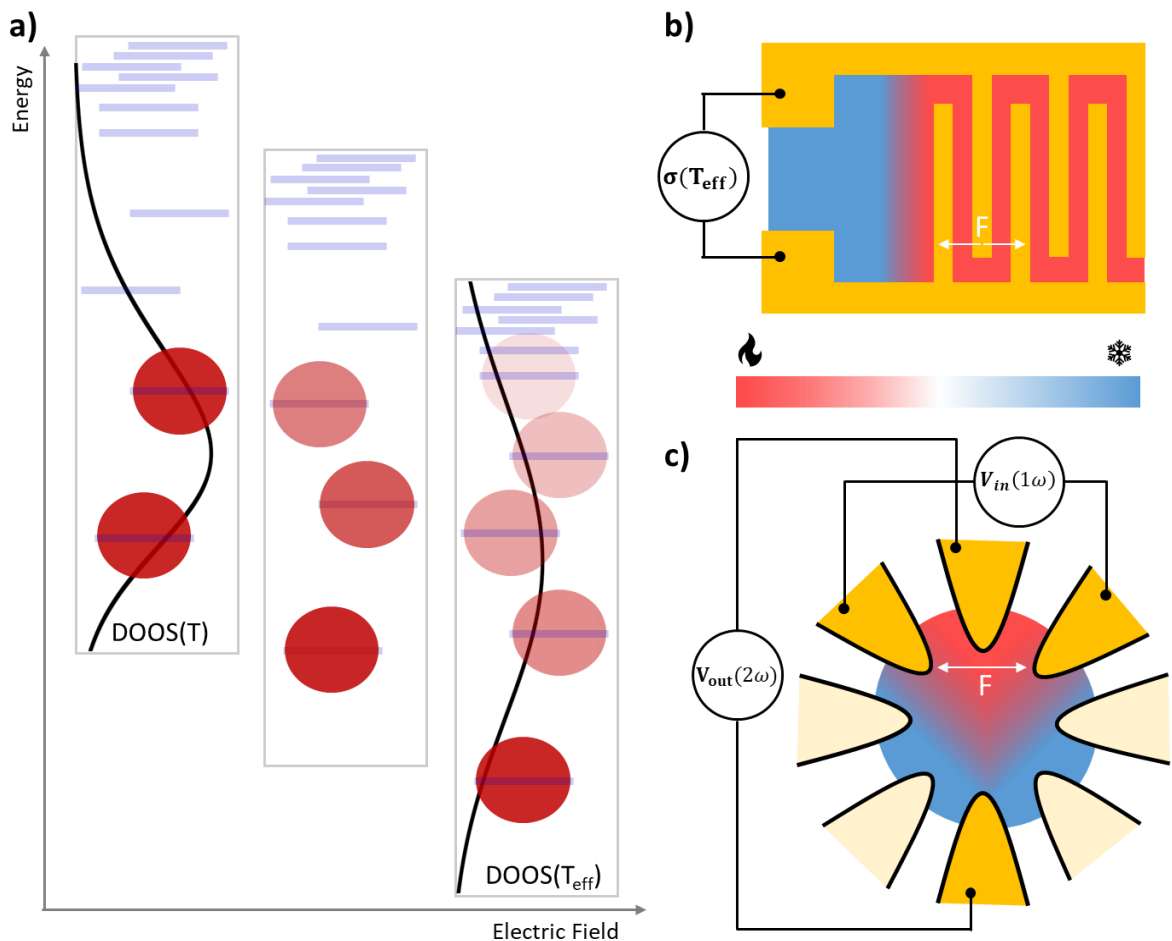

**Figure 1.** Schematic depiction of the experimental concepts. a) The application of an electric field shifts the available localized states (blue lines) for charge carriers (red circles, shading represents the occupation probability) hopping in energy. This allows the occupation of additional states, leading to a shift of the density of occupied states (black line) as if the sample had a higher lattice temperature. b) Depiction of the indirect conductivity-based experiment. Red (blue) colors depict high (low) field and thus high (low) effective carrier temperatures. c) Depiction of the direct Seebeck-effect-based experiment with same color coding as in b).

To achieve the goals defined above, we selected thin films of zinc oxide (ZnO) quantum dots (QDs) as a representative example of a quantum dot solid. First and foremost, this allows us to extend the effective temperature concept to QD solids. Despite widespread use of these materials in applications including commercial displays and solar cells, and the fact that these materials are strongly disordered, the potential field dependence of the conductivity of QD solids so far received very limited attention.[16–18] The specific choice for ZnO-QD is motivated by that they offer good stability under prolonged measurement stress and have a tunable resistance (through UV illumination). Moreover, ZnO QD solids are considered to be a disordered material where the charge conduction mechanism can be modeled with commonly known models for charge transport in disordered systems.[19–21] To our knowledge, there are no direct and few indirect previous measurements of the effective temperature in quantum dot films. The work on quantum dot arrays by Lin et al. touches on the subject but does not analyze its data in the effective temperature framework.[22]

**Methods and materials**

Direct effective temperature measurements were made on 8-tip Au/Cr measurement structures as schematically shown in fig. 1c, fabricated by UV-photolithography on glass substrates. The indirect measurements were conducted on commercially available interdigitated electrodes with a nominal electrode distance of 5 micrometer. The ZnO QDs used in this study are commercially supplied and caped with butyrate ligands and were used as received. For thin film fabrication, a solution of QDs at a concentration of 2 mg/mL in DMSO was drop-casted on both interdigitated and 8-tip structures and dried at 90 °C in air. All experiments are performed in a high-vacuum probe station. Full experimental details are given in Supporting Information (SI) section 1.

ZnO is known to be an inherently n-type material due to the abundance of oxygen defects,[23] while its conductivity depends on the concentration of the atmospheric hydroxyl (OH) and oxygen ($O_2$) species attached to its surface. These surface adsorbates trap electrons and in the case of QDs create a depletion shell, which reduces the effective diameter of the conductive (un-depleted) ZnO core.[21] In previous works, we argued that the energetic disorder in such systems is mainly dominated by the size distribution of the QDs.[19] Hence, by controlling the surface OH and $O_2$ concentration through UV-illumination, we can systematically change the distribution of the effective diameter in our systems which in turn changes the energetic disorder of the system.[19] In SI section 2, a more detailed account is given of the current understanding of the photochemistry at ZnO surfaces. In either case, the UV-tunable effective diameter makes the ZnO QDs an ideal material system for comparing the direct and indirect methods of effective temperature measurements as it allows us to perform multiple measurements on the same sample by systematically changing the material properties without having to rely on multiple synthesis runs. We performed measurements at different OH/$O_2$ concentrations by exposing the sample to UV light, reducing the (surface) OH/$O_2$ concentration, and by keeping the sample at room temperature, allowing it to gradually recover, increasing the (surface) OH/$O_2$ concentration.[24–26]

**Results and discussion**

For the indirect conductivity-based measurements of the effective temperature, one needs to measure the temperature- and electric field-dependence of the conductivity. To determine the temperature dependence of the conductivity, the sample was first cooled to 78 K, and a temperature sweep was conducted in 2 K increments from 80 K to room temperature. At each step, we waited approximately 15 minutes to assure that the sample had reached thermal equilibrium. Fig. 2a shows the electrical conductivity of the ZnO QDs at different OH concentration levels. As expected, the lower the OH concentration is, the higher the conductivity is. The solid lines represent the fits to the common expression for the stretched exponential temperature-dependence of hopping transport in disordered systems:

$$\sigma(T, F \sim 0) = \sigma_0 \exp\left(-\left(\frac{T_0}{T}\right)^\alpha\right) \qquad (3)$$

Here, the parameters $\sigma_0$, $T_0$ and $\alpha$ are free fitting parameters. The temperature exponent in this case is a measure of the relative width of the size distribution of the QDs, that increases when mean diameter decreases with increasing OH-recovery (UV on $\rightarrow$ UV off3), in turn leading to the exponent $\alpha$ getting smaller.[19]

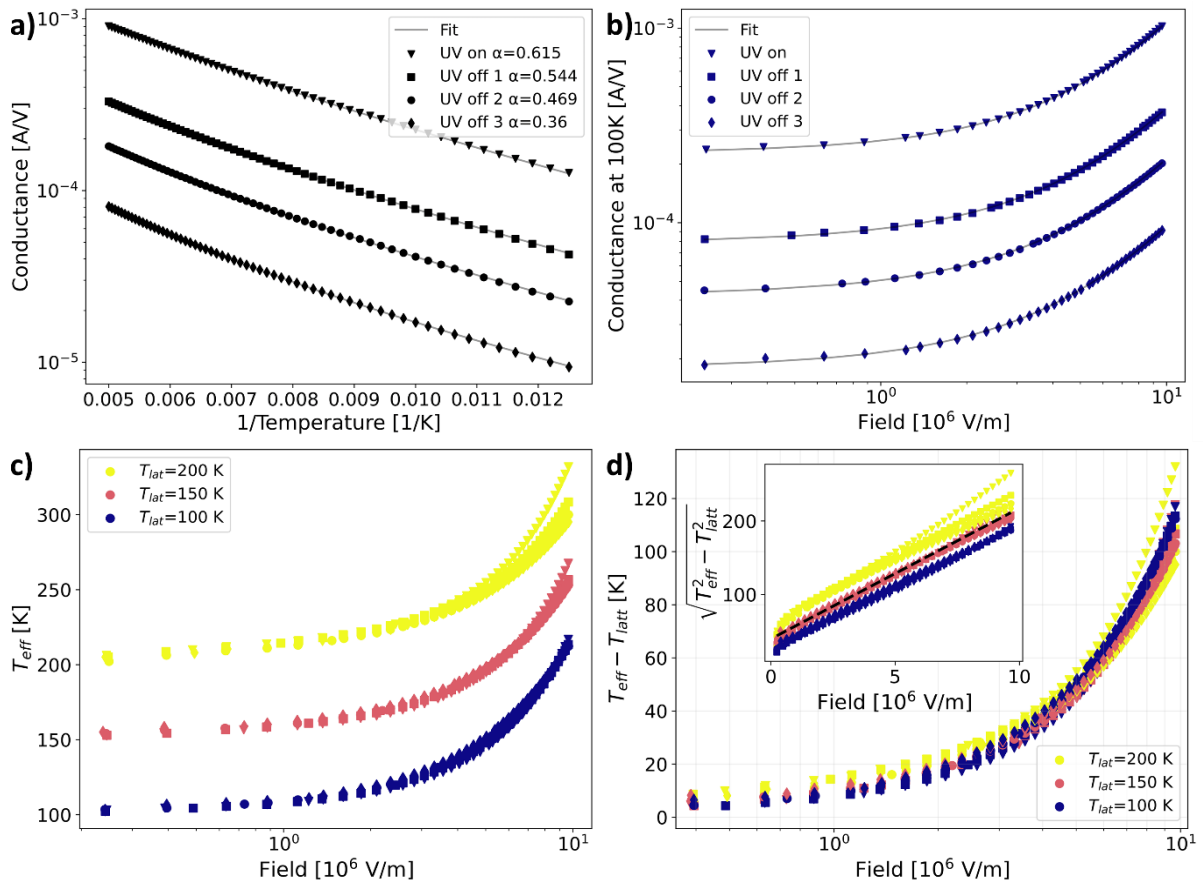

**Figure 2.** a) Temperature-dependent conductance of butyrate-coated ZnO QDs at different OH concentration levels. Lines are fits to eq. 3 with exponents specified in the legend. b) Field-dependent conductance at 100 K for the same sample fitted with a quadratic function to guide the eye. c) Effective temperature at different lattice temperatures. The different markers indicate different levels of hydroxyl exposure as in panels a and b. d) Difference between lattice and effective temperature. Inset: Calculation of the field term in eq. 1, showing that all lattice temperatures and hydroxyl states share a common slope, corresponding to a localization length $\alpha_l = 1.72 \pm 0.13$ nm, as indicated by the dashed line. Error bars for individual measurements are smaller than the symbol size.

Once the temperature dependence of the ohmic conductivity of the system is known, we perform the high electric field conductivity measurements at selected lattice temperatures of 100, 150 and 200K, see fig. 2b for one of them (100 K). By rearranging eq. 3 to obtain eq. 4 and knowing the fit parameters from the temperature dependence, we insert the measured field-dependent conductivities $\sigma(T_{lat}, F)$ to calculate the effective temperature. As explained, this method assumes that the field-dependent conductivity scales in the same way as the temperature-dependent conductivity. The results can be seen in fig. 2c.

$$T_{eff}(T_{lat}, F) = \frac{T_0}{\log\left(\frac{\sigma(T_{lat}, F)}{\sigma_0}\right)^{\frac{1}{\alpha}}} \quad (4)$$

At lower fields, the sample is in the ohmic regime where the effective temperature is independent of the field and equal to the lattice temperature. As the field increases, the conductivity increases which results in the increase of the effective temperature. Subtracting the lattice temperature from the effective temperature (fig. 3d) enables a direct comparison of the functional form and magnitude of the induced effective temperature increase at different lattice temperatures and OH concentrations. We note that the functional form is essentially the same for all UV levels ($\propto \alpha F$ according to eq. 1) without systematic deviations in slope. While we do not have a full explanation for the minor temperature-dependent offsets visible in the inset, we suspect them to be an artefact caused by a small but finite contact resistance that affects the low-field region. The overall consistency in $\Delta T_{eff}$ supports the robustness of the indirect method but, as discussed above, it is no proof that $T_{eff}$ is a real electron temperature and not a mere phenomenological parameter.

The direct, Seebeck effect-based effective temperature measurement follows the principles described in Ref. [15]. The experiment is driven and measured by a Zurich Instruments MFLI lock-in amplifier. Fig. 1c is a schematic of the indirect measurement setup. To provide flexibility to select the optimal measurement geometry, we opted for an 8-tip sample design as illustrated in fig. 1c and shown in detail in SI fig. S1). We established the optimal driving and probing geometry as one where three adjacent electrodes in the 8-tip sample are used as driving electrodes. The middle one is grounded while the outer ones are connected to the differential voltage outputs of the lock-in amplifier. We apply a 10Hz AC signal to avoid any RC-cutoffs, cf. fig. S2. This generates an oscillating field and therefore, potentially, an enhanced effective temperature with double the frequency between the input tips. The tip opposite the grounded tip is connected to the lock-in amplifier's input and hence measures the Seebeck voltage generated due to any (effective) electronic temperature gradients between the driving and the measuring tips. All other terminals of the 8-tip sample are kept floating. Measuring in other configurations such as utilizing two adjacent tips for driving gave qualitatively similar results. As mentioned in the introduction, the advantage of using an AC signal is that the signal of interest will have double the frequency of the driving voltage which enables us to filter for the second harmonic of the input signal via the lock in amplifier.

Since the geometry is complex, we will need to account for some particularities in post-processing to get the true effective temperature from the measured Seebeck signal. In particular, we must correct for the sample resistance being of the same order of magnitude as the input impedance of the lock-in amplifier and the presence of finite fields (and potentially enhanced effective temperatures) between all tips, including the probe tip. Details are given in the SI, section 3.

Seebeck voltages after these corrections are depicted in fig. 3c. The corresponding difference between the lattice and the effective temperature, cf. fig. 3d, is then calculated by dividing the (corrected) voltage by the Seebeck coefficient of the material. To perform that step, that is, to correctly translate measured Seebeck voltages into effective temperatures, we performed additional measurements to determine the temperature-, OH concentration- and lattice temperature-dependence of the Seebeck coefficient of the same ZnO QDs on a separate sample, using a DC Seebeck experiment as described by Zuo et al.[27] Since this necessarily requires a different setup and sample geometry and since the recovery of the OH-surface states varies between runs, we needed a method to transfer the calibration data in fig. 3a to our actual 8-tip samples. To this end, we assume that the Seebeck coefficient directly after UV exposure is identical in all geometries and subsequently, we scale $S$ according to the relative resistance change, cf. fig. 3b.

Although a full discussion of the temperature- and OH-concentration dependence of the Seebeck coefficient of our samples is beyond the scope of the present work, we notice that an increasing $|S|$ with temperature has been reported before for ZnO.[28] Likewise, the increase in $|S|$ with sample resistivity is in line with what one typically observes for disordered semiconductors.[29] The –entirely empirical– linear fits in fig. 3a, b are used in the conversion of the field-dependent Seebeck voltages to effective temperatures: to account for the changing effective temperature of the sample and the correspondingly changing Seebeck coefficient, we iteratively calculate updated effective electronic temperatures from the value for the Seebeck coefficient that corresponds to the current (estimated) value of $T_{eff}$ for a given sample resistance (red crosses in fig. 3b). The procedure is initiated using $T_{lat}$ as initial guess for $T_{eff}$ and converges in few iterations.

After accounting for the (effective) temperature- and therefore field-dependent Seebeck coefficient, all measurements at fields up till ~4×10$^6$ V/m show a very similar behavior that is qualitatively and quantitatively consistent with the indirect measurements in fig. 2d. We attribute the minor differences in functional form to imperfections of the correction method outlined above and field inhomogeneity – we estimate the mean driving field as voltage over electrode distance. For larger fields and lower lattice temperatures, the curves deviate more strongly from each other and no longer follow the functional form eq. 1. Although we cannot formally rule out that these deviations indicate a breakdown of the effective temperature concept, i.e. eq. 2, the fact that the curves for higher lattice temperatures continue to follow eq. 1 suggests they rather reflect an artefact of the sample geometry that leads to an inhomogeneous field (and hence $T_{eff}$) distribution that our simplified model does not accurately account for.

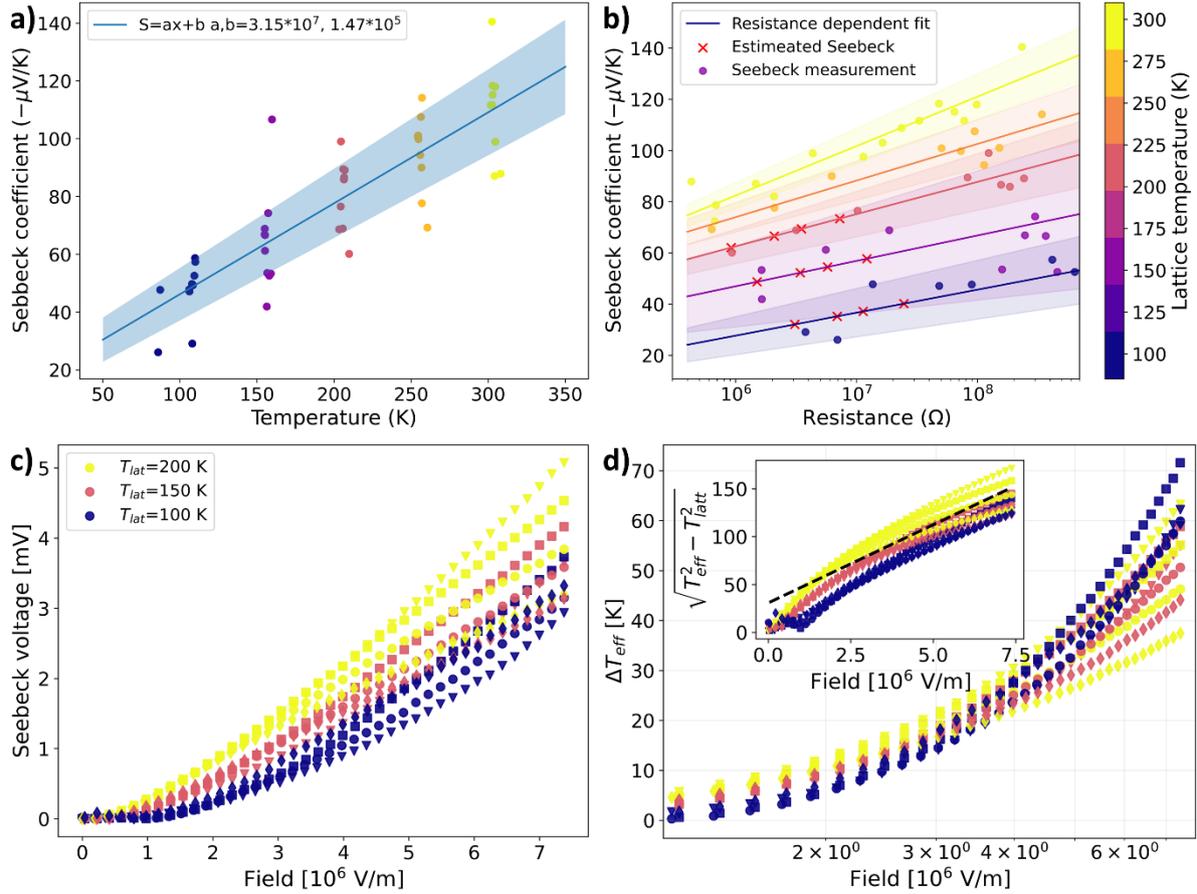

**Figure 3.** a) Absolute value of the Seebeck coefficient measured for different lattice temperatures (color coding) and surface state conditions (UV level) of the ZnO QD solid. The temperature behavior was fitted via a linear increase shown as a blue line with the shaded area demarking the one sigma uncertainty of the fit. b) Same data as in a) plotted over sample resistance with phenomenological logarithmic fits describing the resistance scaling of the Seebeck coefficient for each temperature. Red crosses indicate the estimated Seebeck values utilized in the conversion of Seebeck voltages (panel c) to effective temperatures (panel d). c) Seebeck voltages measured via the direct scheme, dependent on the surface state of the QDs. The different markers represent variations in surface state as in fig. 2. d) Same data as in c), converted into temperature differences via the Seebeck coefficient values estimated from sample resistance (red crosses in b) and the Seebeck value scaling fitted over temperature. Inset: Calculation of the field term in eq. 1 with $T_{eff} = \Delta T_{eff} + T_{lat}$. The high-field slope corresponds to $\alpha_l =$ 1.62 ± 0.21 nm, as indicated by the dashed black line.

As shown in fig. 4, both the qualitative trends and the absolute values in the direct and indirect experiments agree reasonably well, even if not exactly. Since structure sizes can deviate by 0.1-0.2 µm we expect some errors in the estimation of the fields – that are anyhow inhomogeneous in the Seebeck-setup. In combination with expected minor Seebeck coefficient estimation errors, the given deviations are well within reasonable limits, allowing us to conclude with relative certainty that the two methods yield consistent results. In particular, the high-field slopes in the insets of figs. 2d, 3d and 4 around 1.6 nm are the same for both experiments well within the one sigma error margin of the experiments. We therefore conclude that the effective temperatures, as extracted by the indirect method using eq. 2 indeed reflect a true electronic temperature. In view of the complicated nature of the direct measurement, the indirect method is by far preferred for extraction of, e.g., a localization length. Although a full discussion of the value found here is beyond the scope of the present work, we note that a value around 1.6 nm is well in the range of what is reported in the (limited) literature on the topic.[5,6,30] Moreover, because of the core-(depletion)shell structure of our ZnO QDs, the value should likely be interpreted as an effective localization length. [31]

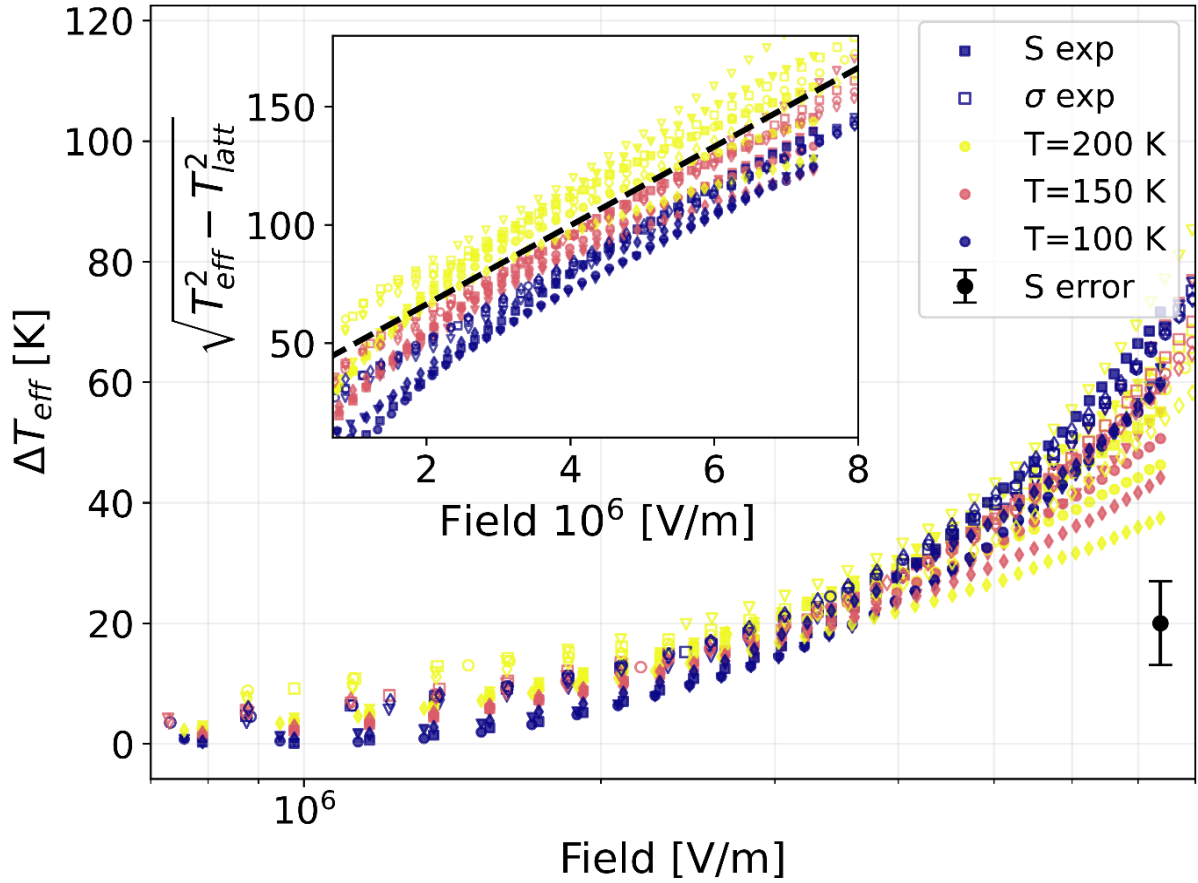

**Figure 4.** Comparison of the direct Seebeck experiment (full symbols) and the indirect conductivity experiment (open symbols), adjusted to the temperature differences of the former as explained in the SI, section 4. Differently shaped markers represent changes in surface state as in fig. 2 and 3. The inset shows the field term in eq. 1. The high-field slope corresponds to $\alpha_l = 1.62 \pm 0.21$ nm, as indicated by the dashed black line. The black error symbol indicates the estimated typical statistical error in the direct Seebeck measurements and mainly results from the fitting error of the Seebeck coefficient estimation, cf. fig. 3.

**Conclusion**

We have performed effective temperature measurements in ZnO quantum dot solids under the application of moderate to high electric field strengths. By comparing by two separate experiments, we could univocally demonstrate that the physical reason for the observed conductivity enhancement is a field-driven shift of the characteristic temperature of the electron distribution (DOOS). Although direct, the Seebeck method requires potentially cumbersome corrections to account for geometrical complexities. The indirect approach, in which the scaling with field is equated to that with temperature, is comparatively easy. While such an analysis was not pursued in this paper, particularly the indirect method opens possibilities to systematically analyze the nature of hopping conduction in these systems in general, by identifying relevant hopping length scales. In general, the demonstration of a true effective electronic temperature in a quantum dot system supports the notion that the effective temperature concept is applicable to a wide range of disordered systems. For the specific system at hand, we note that a significant field dependence of the conductivity already sets in at fields ~$10^6$ V/m that are of the order of, or even below, those found in typical thin film devices like solar cells or light emitting diodes that operate at a several tenths of Volt to several Volt over ~100 nm.

**Conflict of Interest**

The authors declare no conflict of interest.

**Data Availability**

The data supporting this article have been included as part of the manuscript and its Supplementary Information.

## Acknowledgements

We thank Felix Graf for his contributions to the early phases of this project. This work has received funding from the German Research Foundation under Germany`s Excellence Strategy (2082/1 – 390761711). M. K. thanks the Carl Zeiss Foundation for financial support.

Supporting Information to:

**Direct vs. Indirect Measurement of the Effective Electronic Temperature in Quantum Dot Solids**

Anton Kompatscher[+], Morteza Shokrani[+], Johanna Feurstein, Martijn Kemerink[*]

Institute for Molecular Systems Engineering and Advanced Materials, Heidelberg University, Im Neuenheimer Feld 225, 69120 Heidelberg, Germany

[+] contributed equally

[*] corresponding author; email: martijn.kemerink@uni-heidelberg.de

# Contents

# 1 – Experimental details

*Sample fabrication*: The 8-tip measurement structures were realized on glass substrates without the use of any adhesion-promoting treatments. The shape was patterned by a double layer lithography process, using LOR 2A and MAP-1205 as bottom and top layers, respectively. Development was performed in MR-D 526S. Subsequently the metal layers of typically 5 nm chromium and 25 nm gold were deposited using a electron beam evaporator at 10-7 mbar. Liftoff was performed with the remover MR-Rem 700. The samples were cleaned with acetone, isopropanol and demineralized water.

The conductivity-based experiments were conducted on interdigitated electrodes (IDEs) provided by the company Micrux. The company gives a nominal electrode distance of 5 micrometer. From our own measurements on the devices we found distances between 4.1 and 4.3 micron. We calculated with the lower 4.1 number since we expect high field areas to dominate the measurement. (See fig. 5b, d)

The ZnO QDs used in this study are synthesized by NANOXO and caped with butyrate ligands and were used as received. For thin film fabrication, a solution of QDs at a concentration of 2 mg/mL in DMSO was drop-casted on both interdigitated and 8 tip structures and dried at 90 °C in air.

*Experiments*: All experiments are performed in a Janis ST-500-1 vacuum probe station. The pressure during the experiments sits in the $10^{-5}$ mbar range. To ensure reliable thermal contact between the substrate and the mounting chuck, a thin layer of GE varnish (IMI 7031) was applied, and the samples were mechanically secured using the contacting needles.

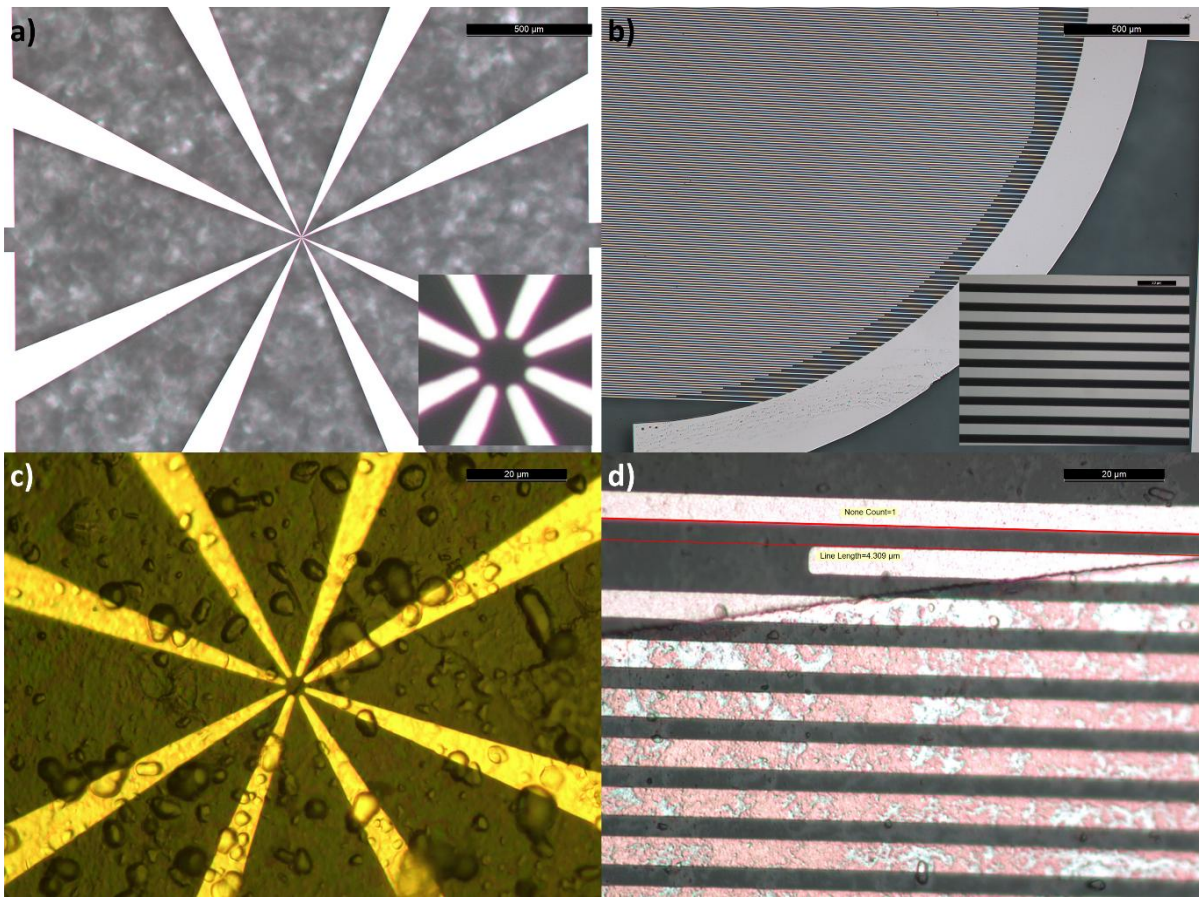

# 1 – Experimental details

**Figure S1.** a) 8 converging tips (bright regions) for the direct, Seebeck-based effective temperature experiments. On the left- and right-hand sides, the edges of larger contact pads are visible. A zoomed-in view of the actual tips is shown in the inset. The diameter of the circular area formed by the tips in the inset is around 4.5 μm. b) Microscope picture of a Micrux interdigitated electrode device covered with quantum dots. c), d) Structures of a), b) covered in quantum dots.

## 2 – Photochemistry of ZnO surfaces

The (unintentional) n-type conductivity of ZnO has been primarily attributed to the existence of oxygen vacancies ($V_O$). Studies have consistently shown that the characteristic green photoluminescence (2.2–2.5 eV) in ZnO, a well-established optical signature for these vacancies, is significantly more intense in materials with a high surface-to-volume ratio, indicating that these optically active defects are predominantly located at or near the surface.[1–4]

This surface proximity creates a complex interplay with the ambient environment. In dark, atmospheric conditions, extrinsic species like oxygen molecules and hydroxyl groups adsorb onto the ZnO surface. Oxygen molecules, in particular, trap free electrons from the conduction band to form negatively charged superoxide ions:

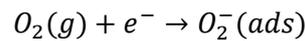
$$O_2(g) + e^- \rightarrow O_2^-(ads)$$

In addition to direct oxygen adsorption, a more complex electron tapping mechanism involving water molecules becomes significant in humid atmospheres. This process involves the dissociative adsorption of water on the ZnO surface.[5–7] The overall reaction can be represented by:[8]

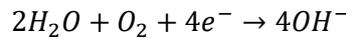
$$2H_2O + O_2 + 4e^- \rightarrow 4OH^-$$

This shows that both adsorbed oxygen and surface hydroxyl groups[9,10] can act as electron traps. Both processes induce a depletion layer around the ZnO QDs which electrically does not contribute to the overall charge conduction and actually hampers it. [11,12]

Upon UV illumination, two simultaneous processes are initiated that increase conductivity. In the first process, photo-desorption of the surface species,[13] the UV light generates a high density of electron-hole pairs. Due to the internal field in the depletion layer, the photogenerated holes migrate to the surface where they recombine with the electrons trapped by the adsorbed oxygen. This neutralizes the superoxide ions, causing them to desorb back into the gas phase as neutral molecules:

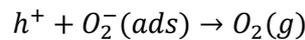
$$h^+ + O_2^-(ads) \rightarrow O_2(g)$$

The resulting increase in conductivity is due to the elimination of these electron-trapping sites and the corresponding shrinking of the depletion layer width. This also leads to a net increase in the concentration of free carriers available for conduction.

In the second process, photo-ionization of oxygen vacancies,[14] the UV energy directly ionizes neutral oxygen vacancies ($V_O^0$) located within the ZnO, causing them to release their trapped electrons into the conduction band and transition into a positively charged state:

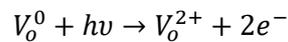
$$V_O^0 + h\nu \rightarrow V_O^{2+} + 2e^-$$

This process serves as a second source of free electrons, further increasing the carrier concentration. This ionization is also the origin of persistent photoconductivity.[15] The change in the vacancy's charge state is accompanied by a significant physical shift of the surrounding lattice atoms, which prevents the immediate recapture of electrons after the UV light is turned off. This causes the enhanced conductivity to persist for an extended period.[14]

Summarizing, oxygen vacancies are the intrinsic source of carriers, while surface species are the (experimentally accessible) extrinsic modulators of the depletion layer. Oxygen vacancies provide the bulk of the free electrons that make ZnO an n-type semiconductor, but they do not, by themselves, directly control the dynamic expansion or shrinking of the surface depletion layer in response to the environment. In contrast, the adsorption and desorption of surface species like $O_2$ and OH groups are the primary mechanisms that directly modulate the carrier concentration *at the surface* by trapping or freeing charge carriers. This dynamic surface process is what governs the thickness of the depletion layer, effectively acting as a gate that controls the overall charge conduction. Moreover, the abundance of charged species in and on the ZnO QDs will contribute to the disordered energetic landscape as experienced by the mobile charges and comes atop the disorder associated with the size dispersion.

We note that the mechanisms outlined above also provide a qualitative framework for interpreting the trends observed in the Seebeck coefficient measurements on our ZnO samples. The photo-ionization of oxygen vacancies and photo-desorption of surface species lead to a substantial increase in the free carrier concentration ($n$). For a degenerate semiconductor like (bulk) ZnO with effective charge carrier mass $m_{eff}$, the Seebeck coefficient ($S$) is inversely related to the carrier concentration, as described by:[16]

$$S = \frac{8\pi^2 k_B^2}{3eh^2} m_{eff} T \left(\frac{\pi}{3n}\right)^{2/3}.$$

While this expression cannot one-on-one be transferred to quantum dot solids, it directly (qualitatively) explains the experimental findings in Fig 3b of the main text. The lowest Seebeck coefficient, which corresponds to the highest conductivity, was measured on the sample state immediately after the UV light was turned off. As explained above, this is the condition where the carrier concentration is expected to be at its maximum due to the combined effects of photo-desorption and the persistent ionization of oxygen vacancies.

## 3 - Frequency dependence

The combination of sample and measurement device resistance and the capacitance of the lock-in amplifier acts as a low pass filter and results in a suppression of the measured signal at higher frequencies. We therefore measured at a frequency of 10 Hz which, as visible in fig. S1, is not affected by this suppression.

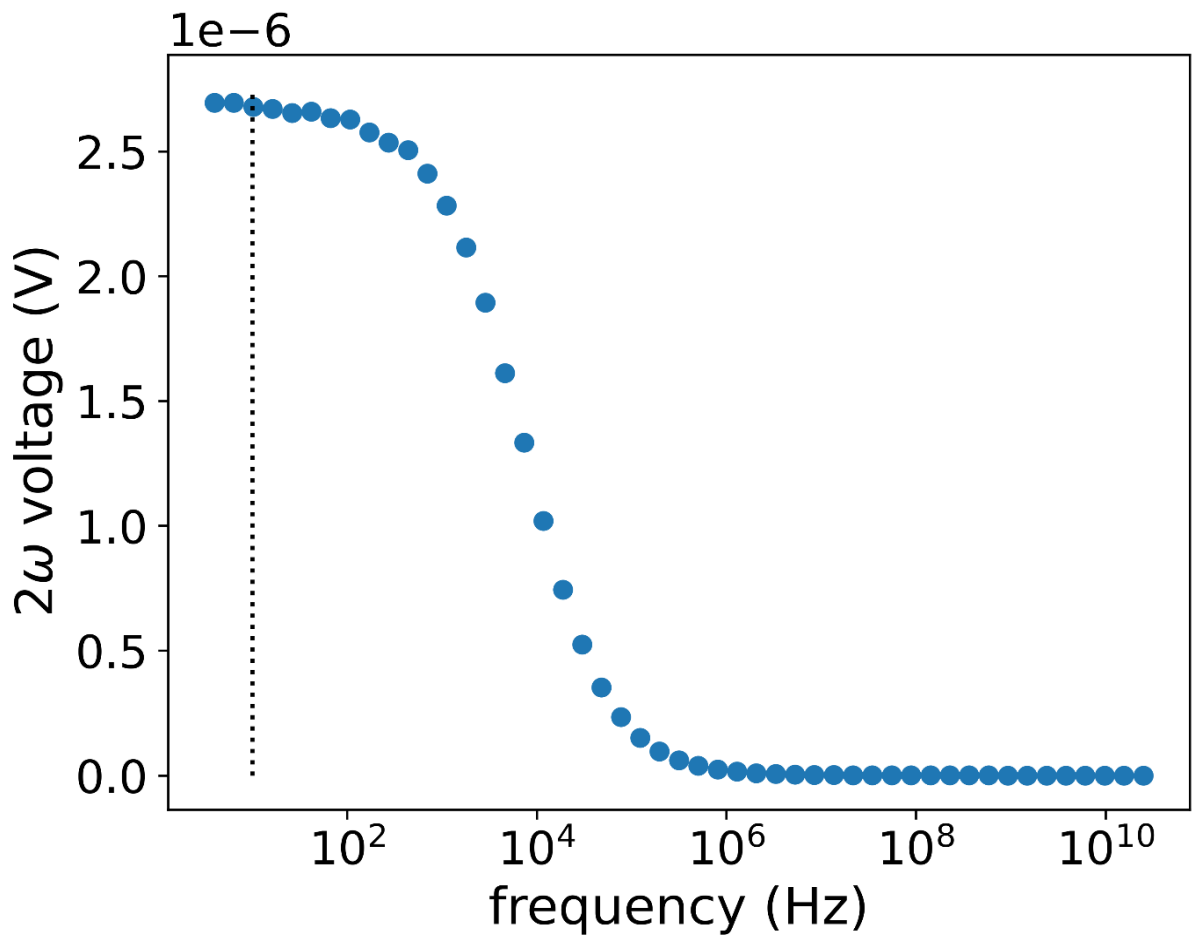

**Figure S2**. Second harmonic output signal of the Seebeck experiment swept over a wide range of frequencies at an input voltage of 20 V. The corresponding sample resistance is 18 MΩ at a temperature of 150 K. The dashed line marks the frequency of 10 Hz at which we measured for our effective temperature experiments.

## 4 - Impedance Corrections

In high resistance samples ($R \geq 1M\Omega$) the impedance mismatch between the 10 M$\Omega$ input impedance of the lock-in and the sample must be corrected for. To this end the connecting resistance (at the relevant field strength) between all relevant tips is measured by a Keithly 2636B source measure unit. The actual generated Seebeck voltage $V_S$ is then calculated from the measured signal $V_M$ as $V_S = V_M * K$ with

$$K = \left( \frac{R_{LI} \| R_{OT1} \| R_{OT2}}{R_{LI} \| R_{OT1} \| R_{OT2} + R_{IO}} + \frac{R_{LI} \| R_{IO} \| R_{OT2}}{R_{LI} \| R_{IO} \| R_{OT2} + R_{OT1}} + \frac{R_{LI} \| R_{OT1} \| R_{IO}}{R_{LI} \| R_{OT1} \| R_{IO} + R_{OT2}} \right)^{-1} \quad (5)$$

the correction factor accounting for part of the overall Seebeck voltage which doesn't drop on the input impedance of the lock-in. The $\|$ symbol indicates the parallel resistance between the components. For this purpose, we have, in lowest order, modeled the rather complicated structure of the system as shown in fig. S3b, where it consists of three parallel equal Seebeck voltage sources accounting for temperature gradients from all input tips to the output tip. Resistances $R$ are labeled after their connections with left- (T1) right- (T2) middle- (I) and output tip (O) as well as the lock-in input impedance. This represents a simplification of the device into three relevant zones which accounts for geometrical complexities in a simplified way. [17] Accounting for the full complexity of geometrically correct addition of all Seebeck contributions scaled by their respective path resistances is not essential to obtain good agreement with the indirect experiment and as such beyond the scope of this publication. Depending on the resistance of the sample, the Seebeck signal $V_S$ is a factor 1 to 15 times bigger than the measured voltage $V_M$.

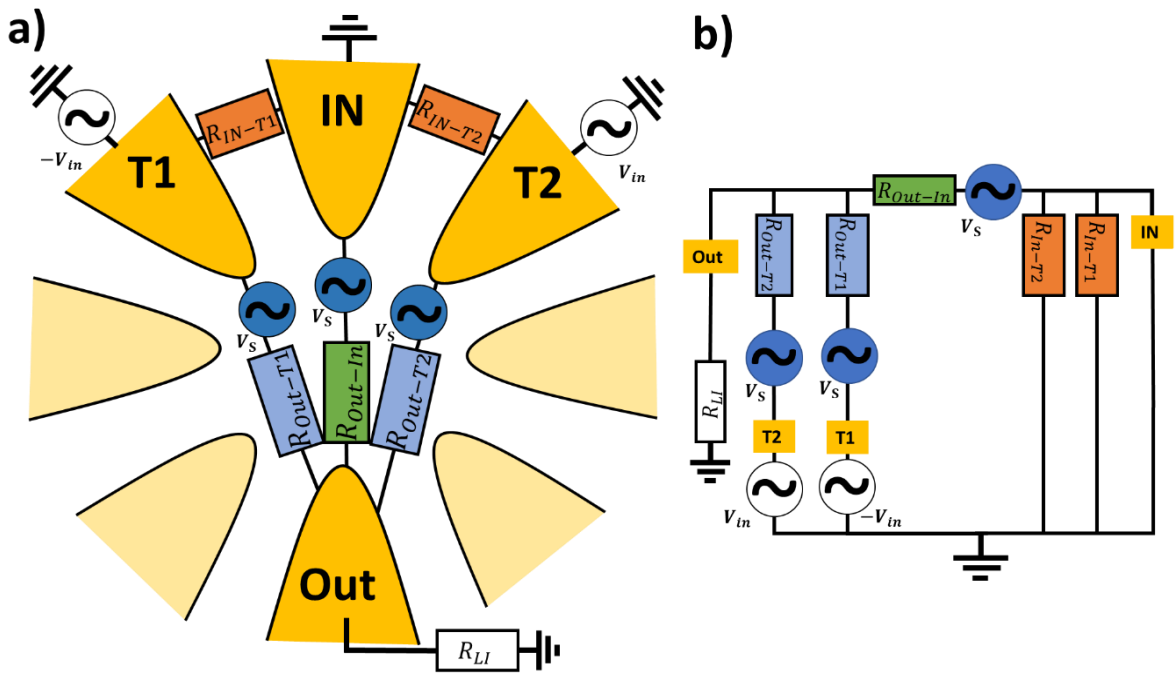

**Figure S3.** a) Schematic depiction of the direct Seebeck measurement structure, including measured resistances and modelled Seebeck voltage sources. The application of an AC driving signal $V_{in}$ results in a Seebeck voltage at double the frequency, modelled as three identical voltage sources $V_S$. The output signal is measured over the MFLI resistance. b) Equivalent circuit of the structure.

## 5 - Adjusted temperature differences

To allow for a quantitative comparison between the direct and indirect measurement schemes, cf. fig. 4 of the main text, we must account for the fact that in the former a non-negligible field exists between the driving (input) and probing (output) tips. This leads to a finite enhancement of the effective electronic temperature at the cold side that is not present in the indirect scheme. Hence, for comparisons we calculate (an estimate for) the effective temperature differences that are present in the direct (Seebeck) experiment geometry, given the temperature scaling measured in the indirect (conductivity) experiment. Specifically, we calculate which field strength would be present between the input and output electrodes of the Seebeck experiment for each given field strength in the conductivity experiment. We then calculate (using the data from the conductivity experiment) the corresponding low-field effective temperature at the probe tip and subtract this from the corresponding high-field temperature at the driving electrodes. To determine the field at the output electrode, we calculate the fraction of the applied voltage that drops over the region between the tips as the ratio of sample resistance to the sample plus measurement resistance. We then divide the voltage drop in the sample by the known distance between input and output tips (~4.36 μm). According to this estimate, the difference in effective temperature between the terminals of the tipped structure is about 20% to 90% of the difference between the lattice and the maximum effective temperature, depending on experimental inputs; the difference quickly shrinks with higher applied fields.